\documentstyle[twocolumn,fleqn,9pt]{article}

\def\thebibliography#1{\section*{References}
 \list
 {[\arabic{enumi}]}{\settowidth\labelwidth{[#1]}\leftmargin\labelwidth
 \advance\leftmargin\labelsep
\setlength{\itemsep}{0ex plus 0.0ex}
\setlength{\parsep}{0ex plus 0.0ex}
 \usecounter{enumi}}
 \def\newblock{\hskip .11em plus .33em minus .07em}
 \sloppy\clubpenalty4000\widowpenalty4000
 \sfcode`\.=1000\relax}

  \advance\oddsidemargin by -1in
  \advance\oddsidemargin by -6mm
  \advance\evensidemargin by -1in
  \advance\evensidemargin by -6mm
  \advance\topmargin  by -1in
  \advance\topmargin  by 3mm
\begin{document}

\twocolumn[


\begin{center}
  {\Large Magnetic-field-induced Luttinger insulator state in
  quasi-one-dimensional conductors}

\bigskip

Victor M. Yakovenko and Anatoley T. Zheleznyak \\
{\bf cond-mat/9807061, July 3, 1998}

\bigskip
Department of Physics and Center for Superconductivity Research,
University of Maryland, College Park, MD 20742, USA\\

\smallskip
\end{center}

{\normalsize\bf Abstract}

   \hspace*{4mm} We present a heuristic, semiphenomenological model of
the anomalous temperature ($T$) dependence of resistivity $\rho_{xx}$
recently observed experimentally in the quasi-one-dimensional (Q1D)
organic conductors of the $\rm(TMTSF)_2X$ family in moderately strong
magnetic fields. We suggest that a Q1D conductor behaves like an
insulator ($d\rho_{xx}/dT<0$), when its effective dimensionality is
one, and like a metal ($d\rho_{xx}/dT>0$), when its effective
dimensionality is greater than one. Applying a magnetic field reduces
the effective dimensionality of the system and switches the
temperature dependence of resistivity between the insulating and
metallic laws depending on the magnitude of the magnetic field and its
orientation along the ``magic'' and ``nonmagic'' angles.


\bigskip\medskip
]

In recent experiments \cite{Jerome95,Jerome96b,Chaikin98}, a very
unusual temperature ($T$) dependence of the longitudinal resistivity
$\rho_{aa}$ was observed in quasi-one-dimensional (Q1D) organic
conductors $\rm(TMTSF)_2ClO_4$ (at the ambient pressure) and
$\rm(TMTSF)_2PF_6$ (at a pressure about 9 kbar) in a moderately strong
magnetic field $H$ of the order of 10 T applied along the {\bf c}
axis.  At high temperatures, $\rho_{aa}$ decreases with decreasing
temperature, as in conventional metals.  However, below a certain
magnetic-field-dependent temperature $T_{\rm min}\sim 20$ K,
resistivity starts to grow with decreasing temperature:
$d\rho_{aa}/dT<0$ at $T<T_{\rm min}$. In other words, the behavior of
the system changes from metallic, $d\rho_{aa}/dT>0$, to insulating,
$d\rho_{aa}/dT<0$, when the temperature is below $T_{\rm min}$.  The
temperature $T_{\rm min}$ increases with the increase of magnetic
field.  As the temperature is lowered further, $\rho_{aa}(T)$
continues to grow until another temperature scale $T_{\rm max}\sim
8\:{\rm K}<T_{\rm min}$ is reached.  At $T<T_{\rm max}$, the behavior
of the system starts to depend crucially on the exact orientation of
the magnetic field \cite{Chaikin98}. If the magnetic field is tilted
to a Lebed's ``magic angle'' \cite{Lebed96}, i.e.\ lies in a plane
formed by the direction {\bf a} of the chains and another integer
crystallographic direction, such as {\bf c} or ${\bf c}+{\bf b}$,
$\rho_{aa}$ recovers the metallic behavior $d\rho_{aa}/dT>0$ at
$T<T_{\rm max}$. For other, generic orientations of the magnetic
field, $\rho_{aa}$ retains the nonmetallic behavior, either continuing
to grow with decreasing temperature: $d\rho_{aa}/dT<0$, or saturating
at a high constant value. The temperature $T_{\rm max}$ does not
depend appreciably on the magnetic field. If at $T<T_{\rm max}$ the
magnetic field is rotated in the plane perpendicular to the direction
{\bf a} of the chains, $\rho_{aa}$ exhibits sharp minima at the
``magic angles''.

This behavior of resistivity completely contradicts the conventional
Fermi-liquid picture of a metal with an open Fermi surface. We suggest
that the following theoretical picture \cite{Yakovenko98d} may
qualitatively account for the unusual behavior of $\rm(TMTSF)_2PF_6$.

The orbital effect of a {\bf c}-axis magnetic field $H$ on a Q1D
conductor is characterized by the cyclotron energy $E_H=ebHv_F/C$,
where $e$ is the electron charge, $b$ is the distance between the
chains in the {\bf b} direction, $v_F$ is the Fermi velocity, and $C$
is the speed of light.  We estimate that $E_H/H\approx1.8$ K/T, so the
magnitude of the cyclotron energy, $E_H\approx14$ K at $H=7.8$ T, is
close to the temperature of the resistivity minimum at that magnetic
field, $T_{\rm min}\sim20$ K.  Taking into account that the minimum of
resistivity clearly has a magnetic origin (it does not exist without
magnetic field), and $T_{\rm min}$ grows with the increase of magnetic
field, we suggest that the minimum of resistivity occurs when the
temperature reaches the energy scale of the magnetic field; that is,
$T_{\rm min}\approx E_H$.

Now we need to identify the nature of the second energy scale in the
problem, the temperature of the resistivity maximum, $T_{\rm max}$. At
the temperatures $T>T_{\rm max}$, it appears that $\rho_{aa}$ depends
only on the magnetic field projection on the ${\bf c}^*$ axis
perpendicular to the {\bf a} and {\bf b} directions. From this
observation, we may conclude that at $T>T_{\rm max}$ the system
behaves effectively as a two-dimensional (2D) system; that is, the
coupling between the chains in the {\bf c} direction is not
relevant. On the other hand, at $T<T_{\rm max}$ the coupling along the
{\bf c} axis becomes important. This is manifested by the magic angles
effect, which is an essentially three-dimensional (3D) phenomenon
involving both the {\bf b} and {\bf c} axes. The coupling between the
chains along the {\bf c} axis is characterized by the electron
tunneling amplitude $t_c$, whose magnitude is believed to be of the
order of 10 K, which is close to $T_{\rm max}\sim8$ K. Thus, we
suggest that the electron tunneling amplitude $t_c$ sets the
temperature scale $T_{\rm max}$ of the resistivity maximum: $T_{\rm
max}\approx t_c$. This conjecture is supported by the experimental
fact that $T_{\rm max}$ (unlike $T_{\rm min}$) does not depend
appreciably on the magnetic field \cite{Jerome96b,Chaikin98}.

Taking into account these energy scales, we identify three qualitatively
different regimes in the behavior of a Q1D system in a magnetic field:

1) High temperatures: $E_H\approx T_{\rm min}<T<t_b$. In
this region, the temperature is greater than both the magnetic energy
$E_H$ and the electron tunneling amplitude $t_c$ along the
{\bf c} axis, but lower than the tunneling amplitude $t_b$ along the
{\bf b} axis. Thus, we may neglect both the magnetic field and the
coupling between the chains along the {\bf c} axis and treat the
system as a normal 2D Fermi liquid without magnetic field. This
results in the quadratic law $\rho_{aa} \sim T^2$ and the metallic
behavior of the resistivity $d\rho_{aa}/dT>0$.

2) Intermediate temperatures: $t_c\approx T_{\rm max}<T<T_{\rm
min}\approx E_H$. In this region, the temperature is still greater
than the coupling between the chains along the {\bf c} axis, so the
system remains 2D; however, the effect of the magnetic field becomes
important. It is known that, in the presence of a magnetic field along
the {\bf c} axis, the motion of electrons along the {\bf b} axis
becomes quantized, and the dispersion law of electrons becomes
one-dimensional (1D) \cite{Lebed84}.  Even though the spectrum of
electrons becomes 1D and their wave functions become localized in the
{\bf b} direction, the wave functions still spread over many chains
(if $E_H\ll t_b$), which results in a considerable interaction between
different chains. So the system is not truly 1D, because it does not
consist of completely decoupled 1D chains. Nevertheless, we may expect
that, at least, some 1D features would be present in this regime and,
via a mechanism that need to be identified, would lead to an
insulating transport behavior $d\rho_{aa}/dT<0$.  The conjecture that
the insulating behavior is caused by the magnetic-field-enforced
``one-dimerization'' requires detailed studying of a specific
mechanism.

3) Low temperatures: $T<T_{\rm max}\approx t_c$. In this region, the
coupling between the $({\bf a},{\bf b})$ planes becomes important. The
magnetic field pointing exactly along the {\bf c} axis does not affect
the electron motion along that axis. Thus, in addition to the
magnetic-field-enforced 1D dispersion law discussed in part 2), the
system acquired an extra dispersion in the {\bf c} direction and
becomes effectively 2D, which results in a metallic, Fermi-liquid
behavior $d\rho_{aa}/dT>0$. If the magnetic field does not point along
the {\bf c} axis, the component of the field perpendicular to the {\bf
c} axis suppresses the energy dispersion along that axis, so the
system remains effectively 1D and insulating: $d\rho_{aa}/dT<0$. If
the direction of the field is close to the {\bf c} axis, we expect
resistivity to decrease with decreasing temperature in the range
$E_H^{(c)}<T<T_{\rm max}\approx t_c$ and start increasing
again at $T<E_H^{(c)}$, where $E_H^{(c)}$ is the
cyclotron energy of the electron motion along the {\bf c} axis, which
is proportional to the projection of the magnetic field perpendicular
to the {\bf c} axis.  The same arguments apply not only to the {\bf c}
axis, but also to the ${\bf c}+{\bf b}$ axis and other integer
crystallographic directions $m{\bf c}+n{\bf b}$. However, because the
electron tunneling amplitudes in these directions decrease rapidly
with the increase of the integers $m$ and $n$, the effect is clearly
visible experimentally only for the ${\bf c}+{\bf b}$ axis.

In summary, we suggest that the unusual transport behavior of
$\rm(TMTSF)_2PF_6$ results from the changes in the effective
dimensionality of the system caused by the applied magnetic field. The
system is 2D at $E_H<T<t_b$ and effectively 1D at
$t_c<T<E_H$. At $T<t_c$ the system is effectively 2D for the
magic orientations of the magnetic field and effectively 1D for
generic orientations. Whenever the system is 2D (or 3D), it is a
normal Fermi liquid, and the temperature dependence of resistivity is
metallic. Whenever the system is effectively 1D, the temperature
dependence of resistivity is insulating.  The latter state of the
system might be called the magnetic-field-induced Luttinger insulator
(MFILI), by analogy with the term ``Luttinger liquid'', which refers
to the metallic state of a 1D system.

In Ref.\ \cite{Yakovenko98d}, we critically analyzed whether various
microscopic models suggested in literature can produce such a behavior
and found that none of the models is fully satisfactory. In
particular, we performed detailed analytical and numerical
calculations within the magnetic-field-induced spin-density-wave
precursor scenario \cite{Gorkov95} and found that the theoretical
results do not agree with the experiment.  Nevertheless, we can
predict some experimental effects based on our heuristic picture. In
Refs.\ \cite{Chaikin94a,Chaikin95}, oscillations of the transverse
resistivity $\rho_{cc}$ upon rotation of a magnetic field in the
$({\bf a},{\bf c})$ plane were observed.  It was found that a small
magnetic field along the {\bf b} axis destroys the oscillations
\cite{Chaikin95}. We predict that, if a magnetic field is rotated in
the magic plane from the ${\bf b}+{\bf c}$ direction toward the {\bf
a} direction, Danner's oscillations should exist, even though the
magnetic field has a finite {\bf b}-component. The suggested geometry
has an advantage over the geometry of the experiment \cite{Chaikin95},
where the magnetic field had a fixed {\bf b}-component, that Danner's
oscillations would not be mixed up with Lebed's oscillations occurring
when the magnetic field is rotated in the $({\bf b},{\bf c})$
plane. Our prediction is based on the idea that Danner's oscillations
require that the electron motion in the third direction is not
suppressed by the magnetic field, which happens only when the magnetic
field belongs to a magic plane.  A two-axes rotation of the magnetic
field was performed in Ref.\ \cite{Naughton98}, but with a different
the choice of geometry.  We also predict that Danner's oscillations
should disappear at $T>T_{\rm max}$, where the electron dispersion in
the third direction is smeared out by temperature.


\begin{thebibliography}{9}

\bibitem{Jerome95} K.~Behnia {\it et al.}, Phys. Rev. Lett. {\bf 74},
5272 (1995).

\bibitem{Jerome96b}
D. J\'{e}rome,  in {\it Correlated Fermions and Transport in Mesoscopic
Systems}, edited by T. Martin, G. Montambaux, and J. Tr\^an Thanh V\^an
(Editions Frontieres, Gif-sur-Yvette, 1996), p. 95.

\bibitem{Chaikin98} E.~I. Chashechkina and P.~M. Chaikin,
Phys. Rev. Lett. {\bf 80}, 2181 (1998).

\bibitem{Lebed96} J. Phys. (Paris) I {\bf 6}, 1819 (1996).

\bibitem{Yakovenko98d} A.~T. Zheleznyak and V.~M. Yakovenko,
cond-mat/9802172.

\bibitem{Lebed84} L.~P. Gor'kov and A.~G. Lebed',
J. Phys. Lett. (Paris) {\bf 45}, L433 (1984).

\bibitem{Gorkov95} L.~P. Gor'kov, Europhys. Lett. {\bf 31}, 49 (1995);
J. Phys. (Paris) I {\bf 6}, 1697 (1996).

\bibitem{Chaikin94a} G.~M. Danner, W. Kang, and P.~M. Chaikin, Phys.
Rev. Lett. {\bf 72}, 3714 (1994).

\bibitem{Chaikin95}
G.~M. Danner and P.~M. Chaikin, Phys. Rev. Lett. {\bf 75}, 4690  (1995).

\bibitem{Naughton98} I.~J. Lee and M.~J. Naughton, Phys. Rev. B {\bf
57}, 7423 (1998).

\end{thebibliography}
\end{document}